# BI-NATIONAL DELAY PATTERN ANALYSIS FOR COMMERCIAL AND PASSENGER VEHICLES AT NIAGARA FRONTIER BORDER

**Zhenhua Zhang**
Department of Civil, Construction, and Environmental Engineering
Institute for Transportation, Iowa State University
Ames, IA, 20010
716-208-0637
Email: zhenhuaz@iastate.com

**Lei Lin**[1]
University at Buffalo, the State University of New York
Buffalo, NY 14260
585-489-2347
E-mail: llin22@buffalo.edu

**Lei Zhu**
The University of Arizona
Tucson, AZ 85721
520-509-5299
Email: leizhu@email.arizona.edu

**Anuj Sharma**
Department of Civil, Construction, and Environmental Engineering
Institute for Transportation, Iowa State University
Ames, IA, 20010
515-294-3624
Email: anujs@iastate.edu

Submitted to TRB 97th Annual Meeting for both Presentation and Publication
July 2017, Washington D.C.

Word Count: 7506
Abstract and Manuscript Text: 5006
Number of Tables and Figures: 3+7=10 (= 2500 words)

---

1 Corresponding Author
This project was partially supported by the National Natural Science Foundation of China (Grant No. 71401012)

**Abstract**
Border crossing delays between New York State and Southern Ontario cause problems like enormous economic loss and massive environmental pollutions. In this area, there are three border-crossing ports: Peace Bridge (PB), Rainbow Bridge (RB) and Lewiston-Queenston Bridge (LQ) at Niagara Frontier border. The goals of this paper are to figure out whether the distributions of bi-national wait times for commercial and passenger vehicles are evenly distributed among the three ports and uncover the hidden significant influential factors that result in the possible insufficient utilization. The historical border wait time data from 7:00 to 21:00 between 08/22/2016 and 06/20/2017 are archived, as well as the corresponding temporal and weather data. For each vehicle type towards each direction, a Decision Tree is built to identify the various border delay patterns over the three bridges. We find that for the passenger vehicles to the USA, the convenient connections between the Canada freeways with USA I-190 by LQ and PB may cause these two bridges more congested than RB, especially when it is a holiday in Canada. For the passenger vehicles in the other bound, RB is much more congested than LQ and PB in some cases, and the visitors to Niagara Falls in the USA in summer may be a reason. For the commercial trucks to the USA, the various delay patterns show PB is always more congested than LQ. "Hour interval" and "weekend" are the most significant factors appearing in all the four Decision Trees. These Decision Trees can help the authorities to make specific routing suggestions when the corresponding conditions are satisfied.

# INTRODUCTION

The border crossing traffic between Buffalo-Niagara Region and Southern Ontario is significant for the economic vitality of both regions. The ability to move goods and people freely and efficiently across the Canadian-US border facilitates the economic growth. A report by the Ontario Chamber of Commerce (OCC) in 2005 puts the value of the annual land-borne merchandise crossing the Niagara Frontier border at $60.3 billion dollars (*1*). Now Buffalo is the second most populous city in New York State and the region of Southern Ontario is also the most densely populated and industrialized region in Canada. Large population and increasing economic activities cause inevitable border-crossing delays and are drawing increasing attentions from both researchers and traffic operators

Buffalo-Niagara Region in Western New York is geographically connected to Southern Ontario, Canada by several bridges over the rivers and canals. There are three main bridges namely the Lewiston-Queenston Bridge (LQ), the Rainbow Bridge (RB), and the Peace Bridge (PB) as shown in FIGURE 1. Due to the increase of continuous travel demand, coupled with tighter security and inspection procedures after September 11, border crossing delay has become a critical problem. As reported by the Ontario Chamber of Commerce, border crossing delay causes an annual loss of approximately $268.45 million for New York State. A press release in 2008 given by Mary E. Peters, the former USA Transportation Secretary, states that the US-bound traffic from Canada encountered delays as high as three hours at several crossings, with delays costing businesses on both Canadian and the US sides as many as 14 billion dollars in 2007 (*2*).

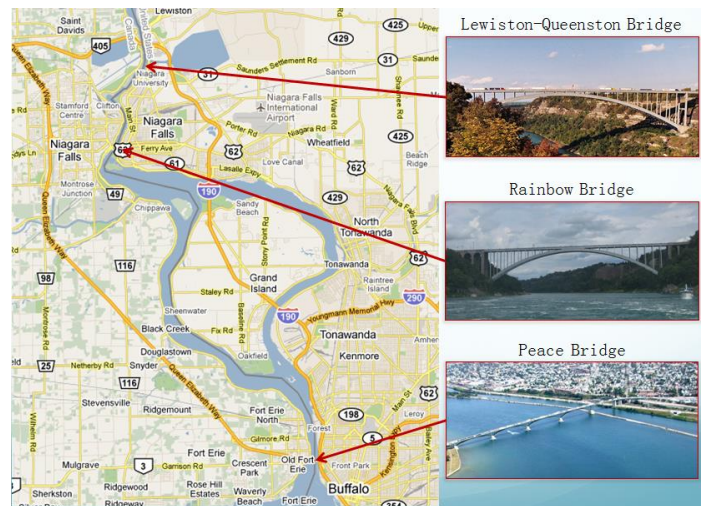

**FIGURE 1 Three bridges at Niagara Frontier border.**

To address these issues, the Niagara International Transportation Technology Coalition (NITTEC), a coalition of fourteen different agencies in Western New York and Southern Ontario, is making a continuous effort in building traffic database and providing real-time border-crossing wait time information to the public. At first, the wait time was obtained based on very rough and approximate estimates of queue length. Later, NITTEC improves the data collection mechanism which uses blue-tooth identification technology to calculate accurate delay. Now, these traffic information are updated every five minutes and can be accessed to motorists through both the official websites and smartphone apps (*3, 4*). Integrated those traffic data into an activity travel demand management (ATDM) system (*5*), the real-time delay data are not only used to help the passengers to plan their trips but also employed in studies for traffic performance measure, flow prediction, etc.

Border-crossing delay problems attract increasing attentions nationwide and there are several insightful studies: Moniruzzaman et al. (*6*) employs the Artificial Neural Network to predict the short-term border-crossing travel time and traffic volume; Maoh et al. (*7*) explores and models the patterns of delays at the three major land crossings connecting Canada to the U.S: Ambassador Bridge, Blue Water Bridge and Peace Bridge and he found that delays at the border vary by season and hour of the day; Gingerich et al. (*8, 9*) found that the total trip time by route is relatively even between US and Canada on Blue Water Bridge and Ambassador Bridge and short-distance journeys had shorter crossing durations on average for trucks; Anderson et al. (*10*) focused the effects of special events and found a significant negative impact of 9/11 can be detected on USA-Canada border crossing. The authors of this paper have been studying border crossing delay problems for more than two years. We have brought the time series concepts in estimating the time-of-day travel pattern (*11*); proposed a two-step delay flow prediction model that consists of a short-term traffic volume prediction model and a queueing model (*11, 12*); In 2015, we even released an Android Smartphone Application for Collecting, Sharing, and Predicting Border Crossing Waiting Time (*13*). Besides these studies, seldom has analyzed the bi-national delay patterns for both commercial and passenger vehicles through the three bridges at Niagara Frontier border.

One can see that most studies are only built and tested based on either passenger or commercial vehicle flow data entering the USA on one bridge. This paper aims to move one step further by exploring the traffic delay patterns on three bridges. Two questions arise:
- For the same vehicle type towards the same bound, are the border delays at the three bridges evenly distributed?
- If not, what factors may lead to the uneven distribution of border delays over the three bridges?

Our yearly efforts have collected enough resources to deliver our research purposes: first, our developed Android Application can filter out useful information from the official website of the Niagara Frontier border crossing authorities (*13*) and collect the border delay data on these three bridges shown in Figure 1; second, another java applications are built to collect the weather data from weather website (*14*) and can be included in the influential factor analysis; besides, the time-of-day, season of the year as well as the special days are also labeled for the study.

Decision Tree Method (*15*) is employed to identify the delay patterns across the three bridges for commercial and passenger vehicles from both the USA to Canada and Canada to the USA. By this method, the significant influential factors can be identified that cause uneven delay distributions. A Decision Tree is a decision support tool that uses a flowchart-like structure in which each internal node represents a "test" on an attribute, each branch represents the outcome of the test, and each leaf node accounts for a class label. It is a modern machine learning model which is simple to understand and to interpret. Based on the Decision Tree findings, the traffic management authority can, therefore, better guide the commercial and passenger vehicles to cross the border and use the limited bridge space more efficiently.

The paper is organized as below. Section 2 introduces the used methodology-Decision Tree to identify various delay patterns based on border-crossing data. Section 3 describes the border wait time data in detail. Model results and insights obtained from these border delay patterns by vehicle type and direction are discussed in Section 4. Finally, the paper ends with conclusions and suggestions for future work.

## METHODOLOGY

Suppose we have a dataset containing $N$ samples. Each sample consists of descriptive features and one target feature. A Decision Tree includes three types of nodes: root node, interior nodes and leaf nodes. The nodes are connected by branches. Each non-leaf node (root and interior) in the tree specifies a test to be carried out on a descriptive feature. Each of the leaf nodes specifies a predicted level of the target feature (*16*). R package *rpart* is utilized to build the Decision Trees in this study (*17*). To grow a Decision Tree, we start at a parent node and split the data on the descriptive feature that results in the largest information gain. The object function is defined in Equation (1) as follows (*18*):

$$IG(D_p, f) = I(D_p) - \sum_{j=1}^{m} \frac{N_j}{N_p} I(D_j) \qquad (1)$$

where, $f$ is the feature to perform the split, $D_p$ and $D_j$ are the dataset of the parent and $jth$ child node, $I$ is our impurity measure, $N_p$ is the total number of samples at the parent node and $N_j$ is the number of samples in the $jth$ child node.

The information gain is the difference between the impurity of the parent node and the sum of the child node impurities. For simplicity and to reduce the combinational search space, each parent node is only split into two child nodes:

$$IG(D_p, f) = I(D_p) - \frac{N_{left}}{N_p} I(D_{left}) - \frac{N_{right}}{N_p} I(D_{right}) \qquad (2)$$

Gini impurity and entropy are two commonly used impurity criteria. *rpart* takes Gini impurity as the default measure. It is defined as a measure to minimize the probability of misclassification:

$$I_G(t) = 1 - \sum_{i=1}^{c} p(i|t)^2 \qquad (3)$$

where, $p(i|t)$ is the proportion of the samples that belongs to class $i$ for a particular node $t$.

The Gini impurity is 0 if all samples at a node belong to the same class, and the Gini impurity is maximal if we have a uniform class distribution. To prevent overfitting which means the Decision Tree is fitting the noise and sample variance in the training set, *rpart* package introduces early stopping criteria into tree induction algorithm. We can stop creating subtrees when the number of samples falls below a threshold, or if the information gain is not sufficient to make partitioning the data worthwhile. In this study, the minimum sample number is set as 100, and the minimum information gain is set as 0.005. As long as either of them is not satisfied, the node will stop splitting.

## DATA FEATURES AND PREPARATION
### DATA DESCRIPTION
A few factors may influence the binational border crossing traffic. The factor name, value examples and data type are listed in TABLE 1.

**TABLE 1 Border Crossing Delay Dataset Summary**

| Feature Category | Name | Value | Data Type |
|---|---|---|---|
| Wait Times | $Bridge_{direction-vehicletype}$ | >= 0 minutes | Continuous |
| Temporal Feature | month | 1,2,…12 | Categorical |
| | season | Spring (month=3,4,5); Summer (month=6,7,8); Fall (month=9,10,11); Winter (month=12,1,2) | Categorical |
| | hour interval | Early_morning (hour=7,8,9); Morning (hour=10,11,12); Afternoon (hour=13,14,15); Evening (hour=16,17,18); Night (hour = 19, 20, 21) | Categorical |
| | weekend | 0-Non-weekend; 1-weekend | Binary |
| | US_holiday | 0-Non-US_holiday; 1-US_holiday | Binary |
| | Canada_holiday | 0-Non-Canada_holiday; 1-Canada_holiday | Binary |
| Weather Feature | temperature | >= 0 °F | Continuous |
| | visibility | 1 (least visible), 2,…,10 (most visible) | Category |
| | precipitation | >= 0 inches | Continuous |
| | weather_condition | Snow, Rain, Clear | Categorical |

In TABLE 1, "$Bridge_{direction-vehicletype}$" represents wait times for a vehicle type (passenger vehicle or commercial vehicle) towards a direction (to USA or to Canada) on a bridge (PB, RB, or LQ). These data range from 7:00 to 21:00 every day between 08/22/2016 and 06/20/2017. For PB and LQ, the wait times are updated every five minutes, 70,404 observations are recorded in total. For RB, the wait times are updated hourly because the Bluetooth technology is not available. We aggregated the five-minute wait times from PB and LQ by taking the average for each hour to make the data on three bridges comparable. In total, 6,850 observations are generated. Based on the Peace Bridge (*3*) and Niagara Falls Bridge Commission websites (*19*), we also find out the number of inspection lanes of the three bridges. As can be seen in TABLE 2, the maximum inpection capacities of PB and RB are almost the same, while there are only 6 inspection lanes for passenger vehicles to USA at LQ. The corresponding temporal features include "month", "season", "hour interval", "weekend" and "holidays of two nations"; the weather features include "temperature", "visibility", "precipitation" and "weather condition". Each hourly wait time is labeled by these features.

**TABLE 2 Inspection Lane Number of Three Bridges**

| Bridge | Inspection Lane Number To U. S. | | Inspection Lane Number To Canada | |
|---|---|---|---|---|
| | Passenger | Commercial | Passenger | Commercial |
| **PB** | 15 | 5 | 11 | 7 |
| **RB** | 16 | N/A | 15 | N/A |
| **LQ** | 6 | 4 | 10 | 5 |

**BORDER-CROSSING DELAY FEATURES**

There are several important empirical findings we can obtain from the raw data. The time-of-day results are very insightful to help us understand the border-crossing problems. Figure 2 selects the time-of-day delay patterns on PB and converts the continuous wait time into discrete delay type: no delay (0 min wait time); slight delay (0 mins < wait times <= 15 mins); delay (15 mins < wait times <= 30 mins); and heavy delay (wait times > 30 mins). The categorization can provide more straightforward traffic delay features. From the figure, one can easily figure out both the time-of-day features and directional features:

- First, the time-of-day features can be seen in passenger and commercial vehicles in both from USA to Canada and from Canada to USA. Even though there is no clear AM or PM peak, one can still see that delays during the daytime are more severe than that during the night.
- Second, on PB, the delays from the USA to Canada are more severe than that from Canada to USA for both passenger and commercial vehicles. This uneven delay features for two directions and its influential factors can be time, weather or other factors.
- Third, commercial vehicles experience overall slighter delays than passenger cars. This may be due to that passenger and commercial vehicles are checked in separate lanes in border crossing.

As the overall delay levels in two directions are different, the passenger or commercial vehicles from the USA to Canada on PB may not return to its original path. It is highly possible that passenger vehicles come to Canada on PB in the morning but return to the USA on RB in the afternoon. If so, one can also see an uneven delay distribution on the three bridges during the same time-of-day. The combination of the delay levels on three bridges and their influential factors are also worth studying.

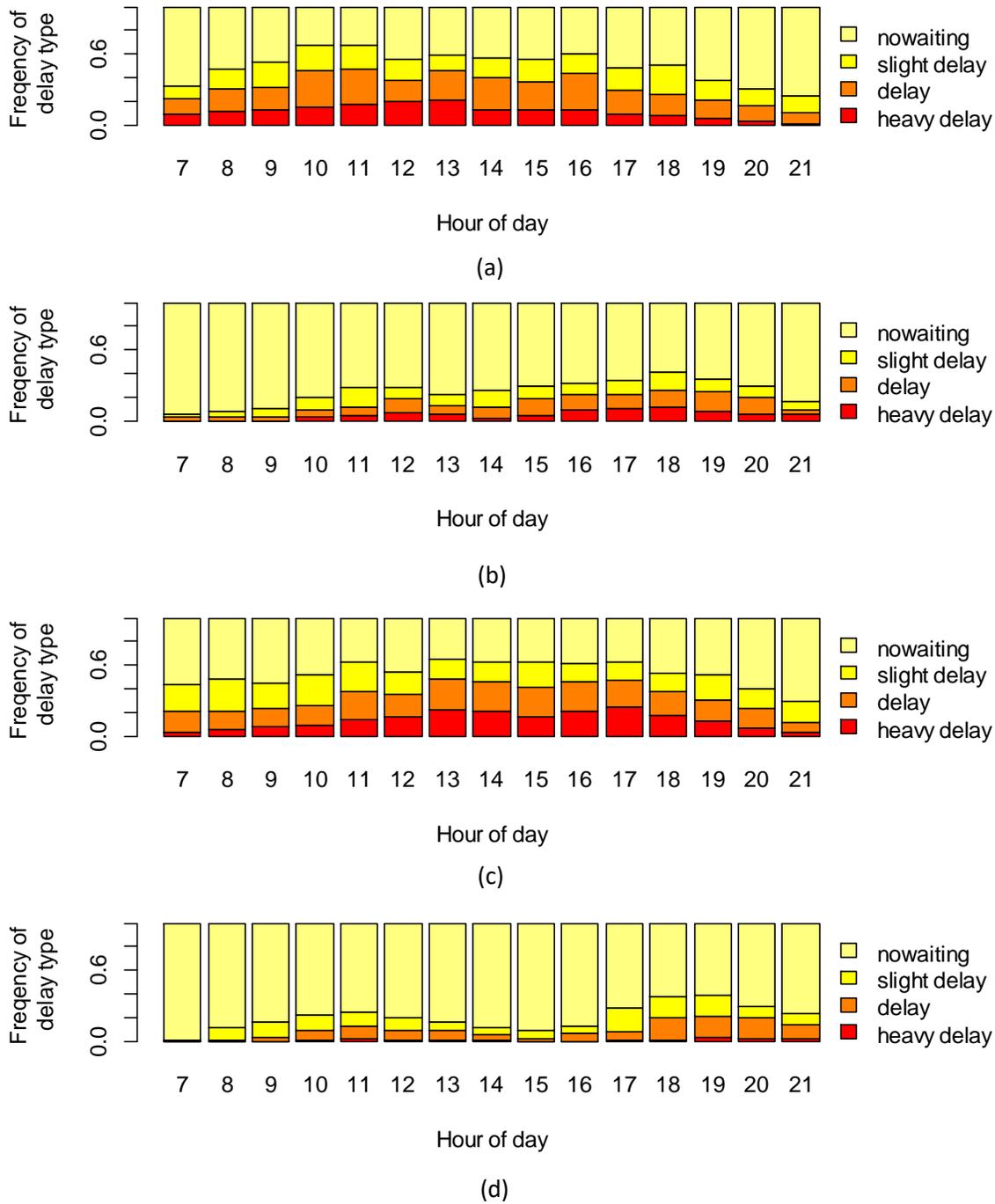

**Figure 2 Delay Type Distribution during Different Hours of Day at PB (a) passenger vehicles from USA to Canada; (b) passenger vehicles from Canada to USA; (c) commercial vehicles from USA to Canada; (d) commercial vehicles from Canada to USA.**

**DATA PREPARATION**
In this sub-section, we prepare the delay data in three bridges of each bound into one single pattern for Decision Tree algorithm. Before applying the algorithm on our data, for each vehicle type towards each direction, we conduct some data preprocessing work:
- The observations are only kept when at least one bridge has non-zero wait times. Instead of situations when all three bridges have zero wait times, we focus on whether the same congestion status occurs across the three bridges at the same time.
- After deleting the records when all three bridges have zero wait times, few zero wait time cases are left for each bridge. Therefore we merge the no delay (wait times = 0 min) and slight delay (0 min < wait times <= 15 mins) into one category as both delay status do not differ much most of the time.
- For each vehicle type towards each direction, concatenate the discretized delay status of each bridge together and form a new feature called delay pattern. "$PB - RB - LQ_{toUS-Pass}$" represents the delay patterns across the three bridges for the passenger vehicles to USA. As there are three delay categories for each bridge, for this new feature, it has 27 different combinations. So is the new feature "$PB - RB - LQ_{toCAN-Pass}$". Note that the trucks are not permitted to go through RB, for "$PB - LQ_{toUS-Truck}$" and "$PB - LQ_{toCAN-Truck}$", there are only 9 possible categories.

FIGURE 3 shows the frequencies of delay patterns across the three bridges for each vehicle type and each direction. As can be seen, for any case the dominant patterns show the delay status across the three bridges is uneven. For example, for $PB - RB - LQ_{toUS-Pass}$, the most frequent pattern is "delay-slight delay-slight-delay". This can briefly answer our first question: the border crossing delays are not evenly distributed over the three bridges for the same vehicle type towards the same bound.

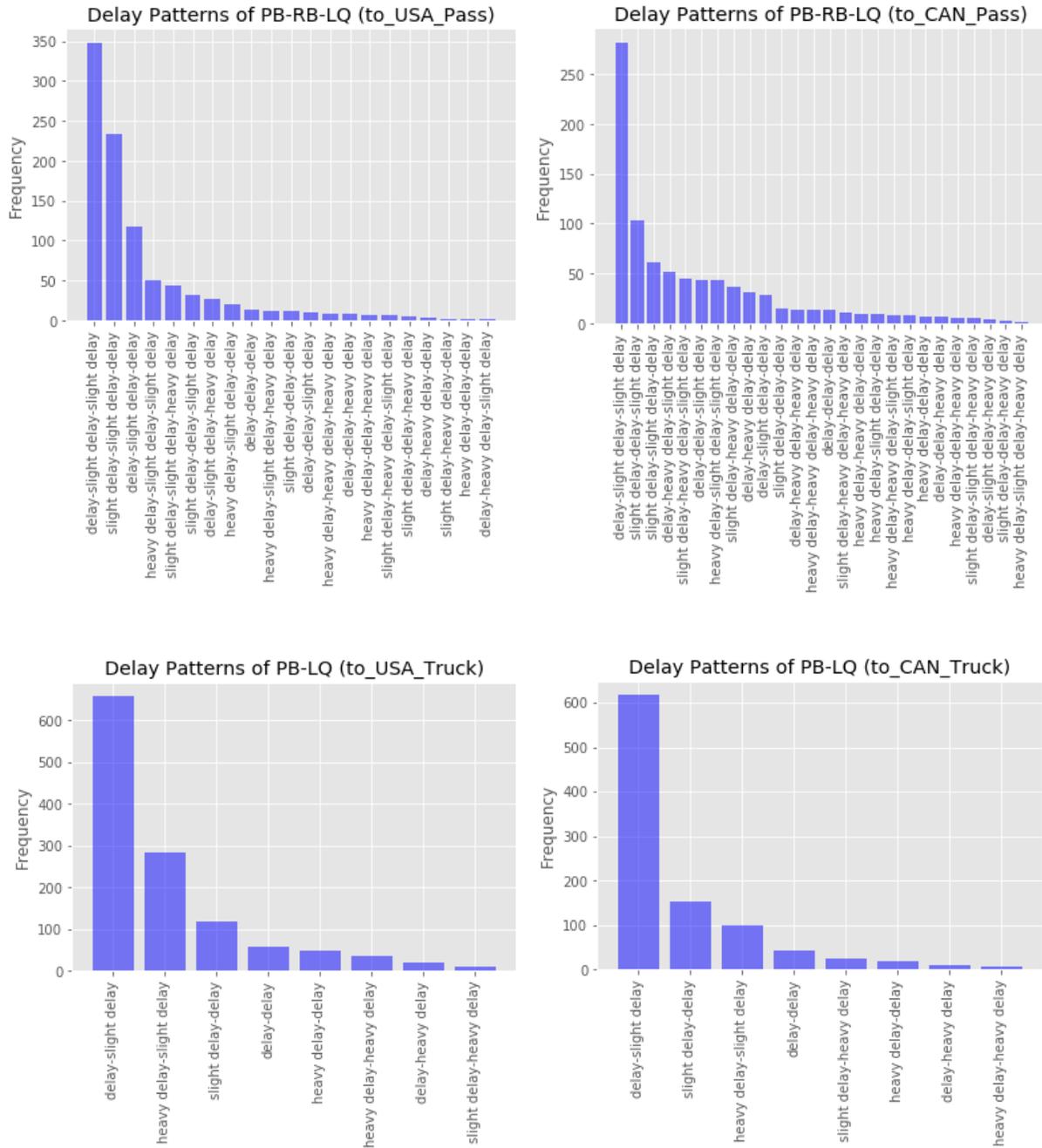

**FIGURE 3 Delay pattern frequencies by vehicle type and direction.**

## DECISION TREE MODEL RESULTS

This section first analyzes the Decision Tree results based on the binational border delay data for passenger vehicles; following that the Decision Trees from commercial truck border delay data are explored; finally, we summarize the findings of four Decision Trees and the corresponding significant factors.

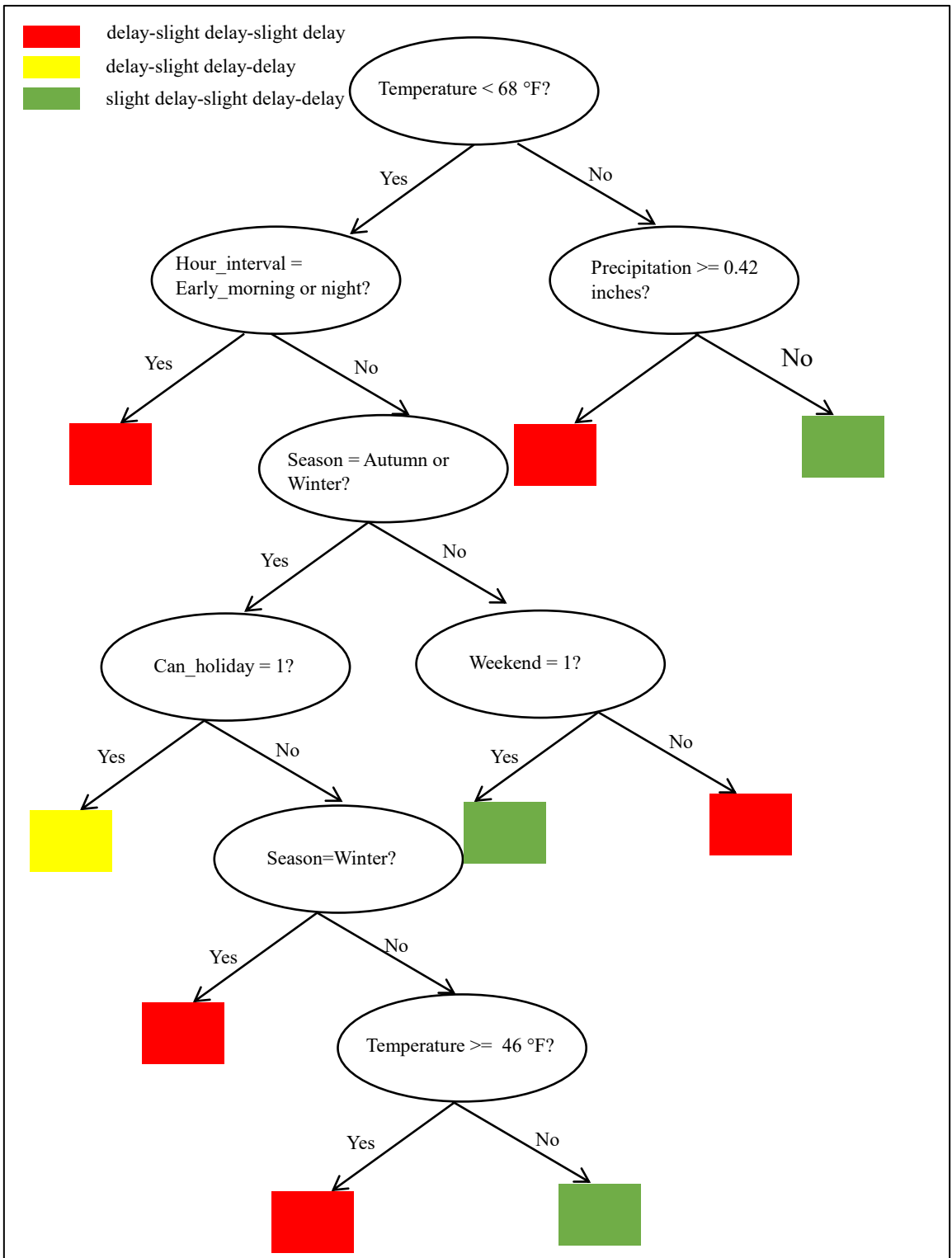

**FIGURE 4 Decision Tree of delay patterns of PB-RB-LQ (toUSA-Pass).**

FIGURE 4 shows the Decision Tree learned by delay patterns of PB-RB-LQ for passenger vehicles to USA. First we can see that the patterns of "delay-slight delay-slight delay" (red rectangle) and "slight delay-slight delay-delay" (green rectangle) are the two dominating delay patterns. These patterns mean that either PB or LQ is a little busier than the other two bridges. The important factors include "hour interval," "season," "weekend," "temperature" and "precipitation." Second, the third delay pattern "delay-slight delay-delay" is possible when "temperature < 68 °F", "hour interval = morning, afternoon or evening," "season = autumn or winter" and "Can holiday = 1". It tells that both PB and LQ can have wait times between 15 mins and 30 mins when the Canadian people come across the border on holiday. However, the RB is still under "slight delay" and more traffic should be re-routed there. Through FIGURE 1, we can see that Bridge LQ connects Ontario Highway 405 in Canada and I-190 in USA; and Bridge PB links Queen Elizabeth Highway in Canada and I-190 in USA. The convenient interchanges between highways may be one reason for the delays in PB and LQ.

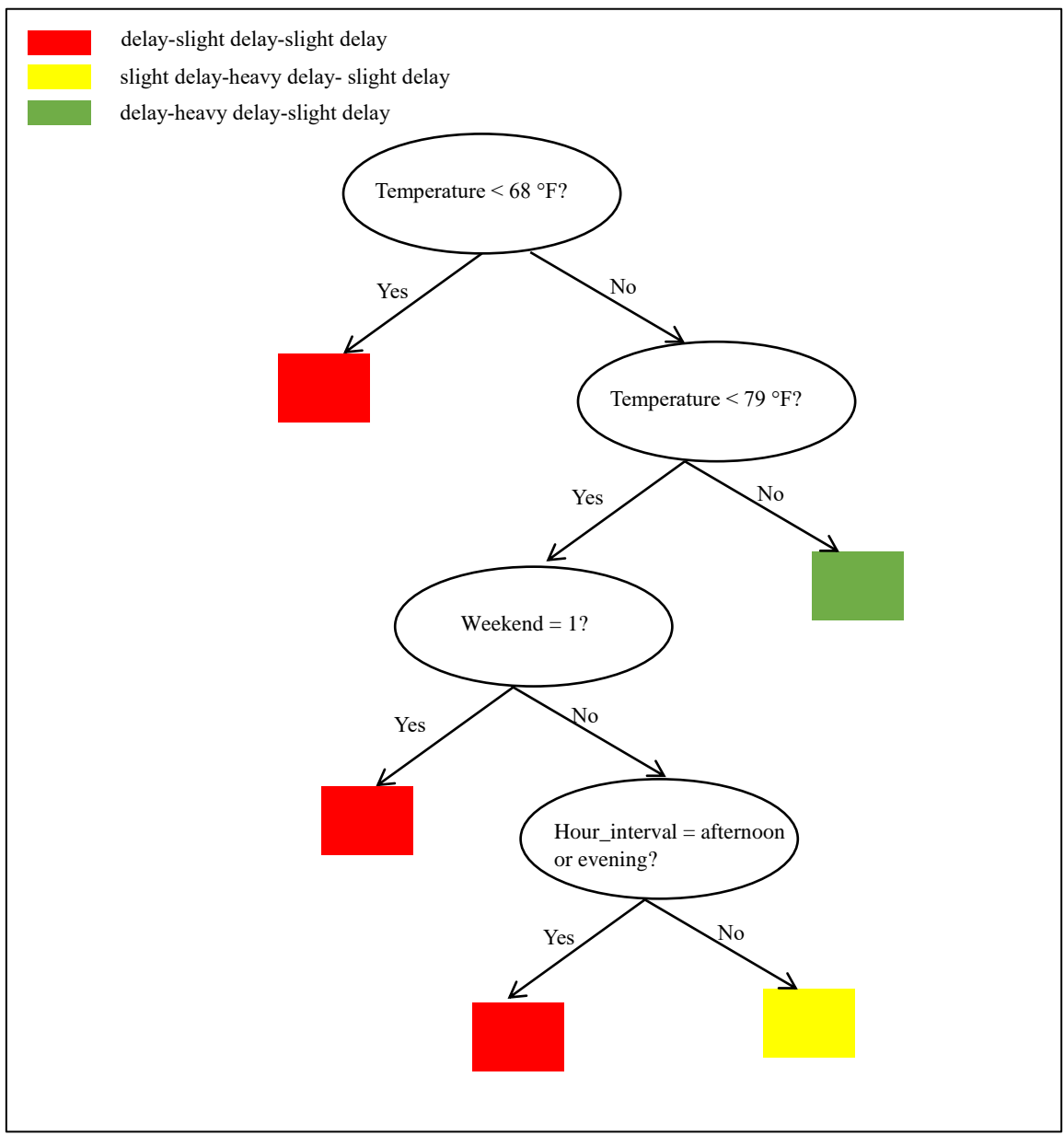

**FIGURE 5 Decision Tree of delay patterns of PB-RB-LQ (toCAN-Pass).**

FIGURE 5 shows the results for passenger vehicles from the USA to Canada, in which the captured delay patterns are different. First, when the temperature is high "temperature >= 79 °F", the "delay-heavy delay-slight delay" pattern (green rectangle) is at the leaf. It means RB is in "heavy delay" status, followed by the "delay" status of PB, while LQ is just in "slight delay." Second, the "slight delay-heavy delay- slight delay" pattern (yellow rectangle) is impacted by factors "temperature," "weekend" and "hour interval." In this pattern, RB is experiencing "heavy delay" status while the other two are just in "slight delay." Third, the left frequent delay pattern is "delay-slight delay-slight delay" for PB-RB-LQ. It shows that PB may also be busier than the other two bridges when some factors "temperature," "weekend" and "hour interval" are satisfied. These patterns show that for passenger vehicles from the USA to Canada, RB may be most congested in some cases such as high temperature. This may be due to that RB is close to Niagara Falls which is a world-famous tourist site. A survey shows that in 2014 there are 8.91 million visitors in total in USA side, including 1.95 million in June, 1.91 million in July and 2.12 million in August, and 6.6 million visitors coming to this area are domestic (*20*). The high volume of visitors may cause the congestion in RB in summer.

After analyzing the border delay Decision Trees for the passenger vehicles over the three bridges, FIGURE 6 shows the delay pattern Decision Tree for commercial vehicles from Canada to USA.  It is worth mentioning that the commercial vehicles are not allowed to go through RB. FIGURE 6 shows that there are two delay patterns uncovered by the Decision Tree. First, the most frequent "delay-slight delay" pattern (red rectangle) appears four times as the leaf node. It may be impacted by "hour interval," "season," "weekend" and "temperature." Second, a more extreme delay pattern "heavy delay-slight delay" (yellow rectangle) also appears twice as the leaf node. One branch is "hour interval=afternoon, evening or night" → "season=autumn" → "weekday" → "temperature >= 64 °F" → "heavy delay-slight delay"; the other branch is "hour interval=afternoon, evening or night" → "season=autumn" → "weekday" → "temperature < 64 °F" → "hour interval=evening" → "heavy delay-slight delay". Both delay patterns show PB is more congested than LQ for commercial trucks to U.S. The reason may again be the convenient location of PB that the commercial trucks can take the Queen Elizabeth Way directly across the PB and proceed on the ramp to I-190 South which connects to the I-90 (*19*).

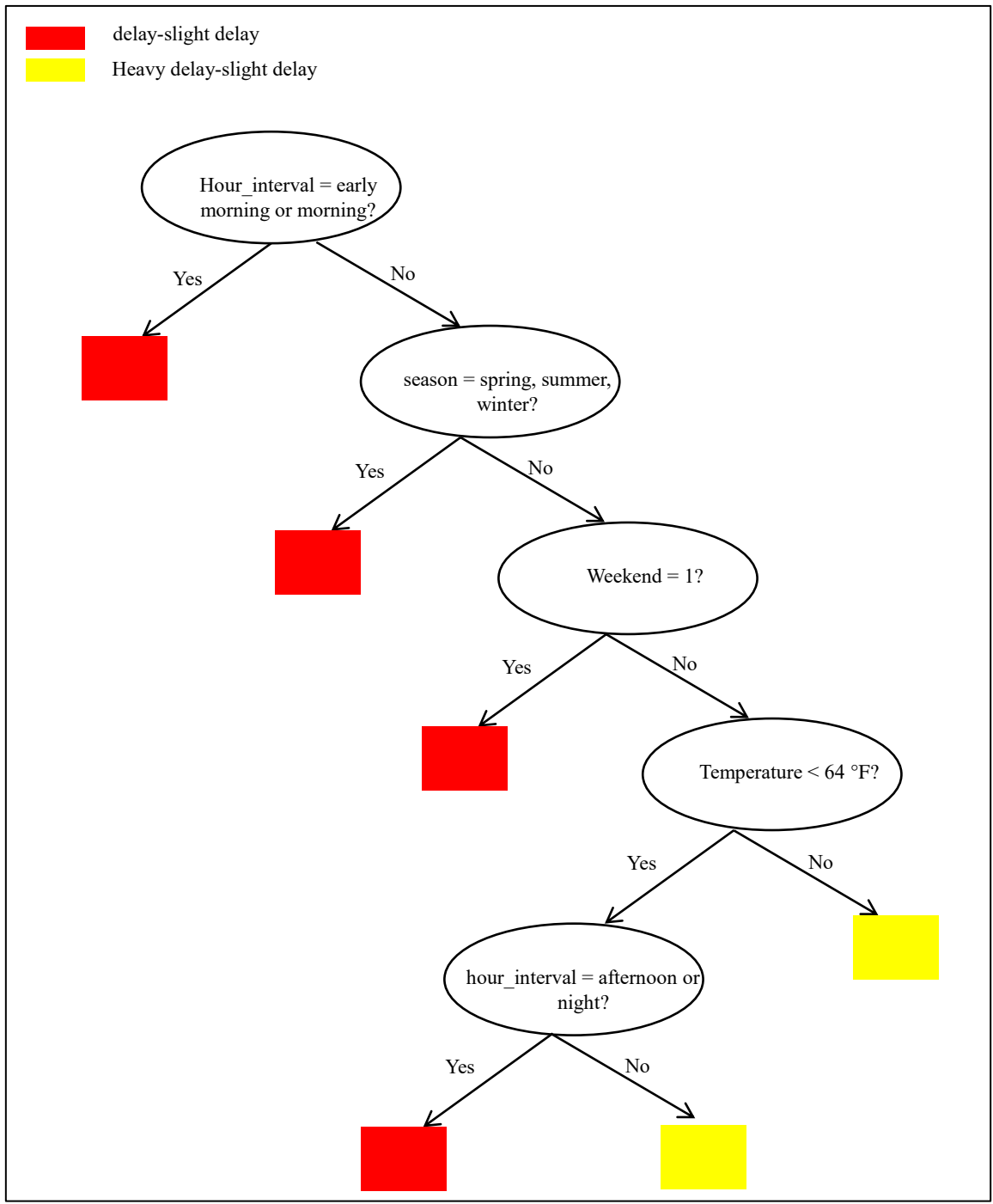

**FIGURE 6 Decision Tree of delay patterns of PB-LQ (toUSA-truck).**

For commercial trucks to Canada, the Decision Tree of delay patterns of PB-LQ is shown in FIGURE 7. There are mainly two delay patterns influenced by "hour interval," "weekend" and "season." Either PB or LQ is in "delay" status, while the other one is in "slight delay" status. No extreme delay pattern is revealed by the Decision Tree like one bridge is in "heavy delay" and the other one is in "slight delay."

Zhang, Lin, Zhu, Sharma                                                                                                15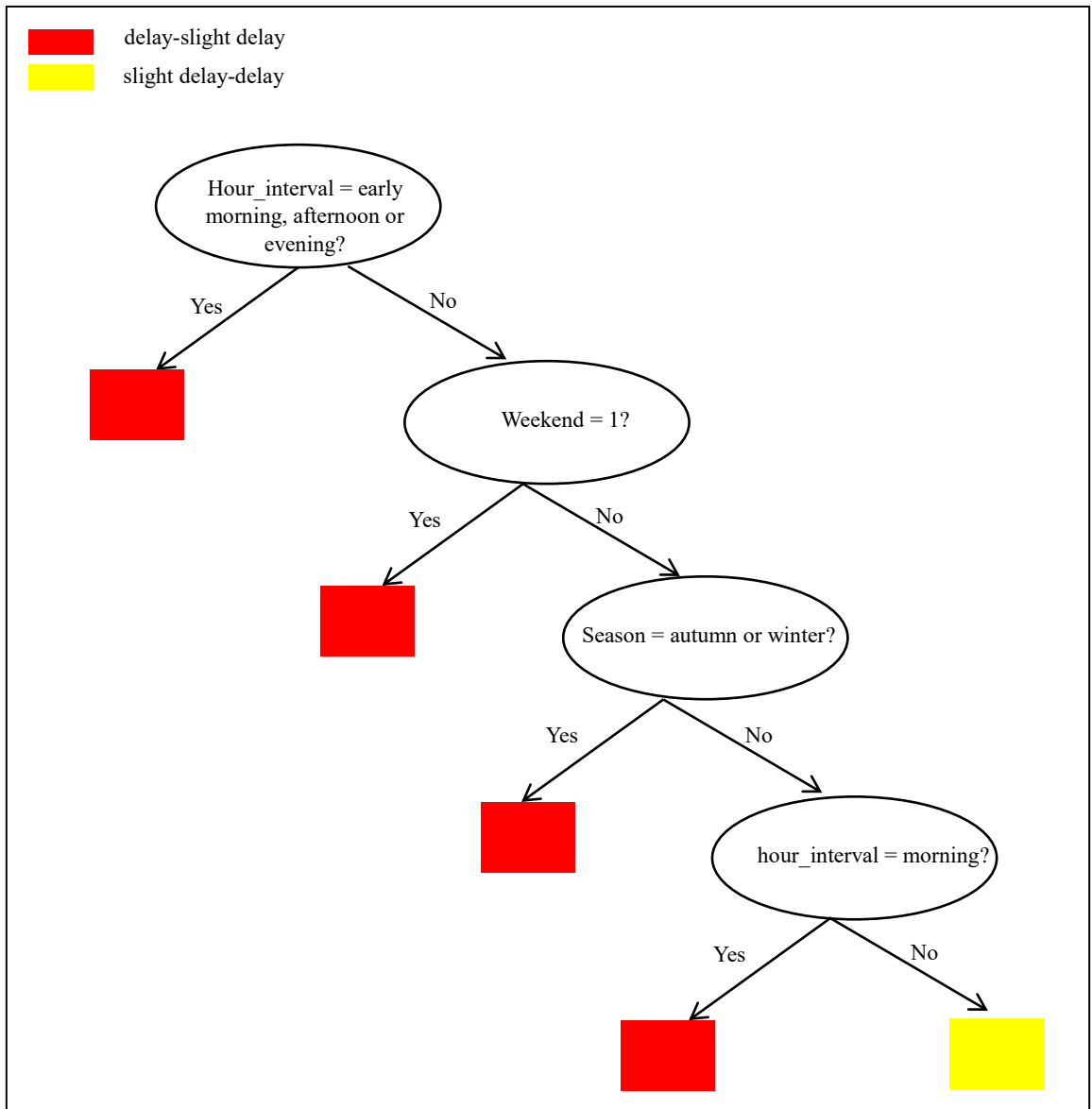

**FIGURE 7 Decision Tree of delay patterns of PB-LQ (toCAN-truck).**

TABLE 3 summarizes the various delay patterns observed at the leaf nodes of Decision Trees, as well as the significant factors at the non-leaf nodes. Several interesting conclusions can be concluded.

First, the Decision Tree results verify that for any vehicle type towards any direction, the border wait times are not evenly distributed based on the dominant frequent delay patterns captured by the Decision Tree. In particular for passenger vehicles to Canada, one bridge (RB) could be in "heavy delay," while the other bridges are in "slight delay." Similarly, for commercial vehicles to the USA, PB could be much busier than LQ;

Second, although the vehicle type is the same, the different directions result in various frequent delay patterns with significant factors. For example, for passenger cars to the USA, RB is always in "slight delay" status, PB and LQ can be in "delay" status impacted by four temporal features including "Can_holiday" and two weather features. For the passenger vehicles to Canada, it is different that LQ is always in "slight delay" status while RB could be "heavy delay." And only three factors are impacting the delay patterns.

Third, in our dataset, we have six temporal features and four weather features. "Hour interval" and "weekend" are included as significant nodes in all the four Decision Trees. "month," "US_holiday" and "weather_condition" never appear in Decision Trees as split nodes.

**TABLE 3 Summary of Delay Patterns by Vehicle Type and Direction**

| Vehicle Type and Direction | Delay Pattern | | | Influential factors |
|---|---|---|---|---|
| | PB | RB | LQ | |
| **Passenger Vehicles to the USA** | delay | slight delay | slight delay | season, hour interval, weekend, Can_holiday, temperature, precipitation |
| | slight delay | slight delay | delay | |
| | delay | slight delay | delay | |
| **Passenger Vehicles to Canada** | delay | slight delay | slight delay | hour interval, weekend, temperature |
| | delay | heavy delay | slight delay | |
| | slight delay | heavy delay | slight delay | |
| **Commercial Vehicles to the USA** | delay | N/A | slight delay | season, hour interval, weekend, temperature |
| | heavy delay | N/A | slight delay | |
| **Commercial Vehicles to Canada** | delay | N/A | slight delay | season, hour interval, weekend |
| | slight delay | N/A | delay | |

## CONCLUSION AND FUTURE STUDY

This paper aims to identify whether the three bi-national border ports (PB, RB, and LQ) for both commercial and passenger vehicles at Niagara Frontier are utilized sufficiently and the significant factors that may result in uneven border crossing delays. A detailed historical border delay data, as well as the corresponding temporal and weather data, are collected. Data preprocessing was conducted to discretize the raw wait time data of the three bridges into delay categories, which are further combined by each vehicle type and each direction. The new delay pattern is taken as the target for the Decision Tree algorithm. Based on the Decision Trees, the most frequent delay patterns by each vehicle type towards each direction are discovered, the significant factors hidden behind are also analyzed. A few key findings are summarized as follows:

- For the passenger vehicles to the USA, the fact that both LQ and PB connect the Canada freeways with USA I-190 may cause the LQ and PB congested, especially when it is a holiday in Canada. For the passenger vehicles in the other bound, the high volume visitors to Niagara Falls in the USA may be a reason why RB is in "heavy delay" in summer.
- For the commercial trucks to the USA, the frequent delay patterns show PB is always more congested than LQ; however, there are no distinct border delay tendencies for the commercial vehicles across the two bridges to Canada.
- "Hour interval" and "weekend" are the most significant factors for both the passenger and commercial vehicles in two border-crossing directions.

In the future studies, more detailed data are necessary. For example, bi-national origin-destination data can help to understand why some bridge is more popular than the others. The total trip duration may be shorter for some travelers although higher border delays are expected. Therefore the authority can make more acceptable route guidance to utilize the border crossing facilities more efficiently.